\documentclass[aps,prd,showpacs,preprint,tightenlines,12pt]{revtex4}

\def \ampq#1#2{#1_{#2}^{q} e^{i \delta^{q}_{#2}}}
\def\sq#1{\sqrt{#1}}
\def\pp{\pi^{+}}
\def\pz{\pi^{0}}

\def\kzbar{\bar{K}^{0}}
\def\km{K^{-}}
\def\sci#1#2{#1\times 10^{#2}}
\def\fr#1#2{\frac{#1}{#2}}
\reversemarginpar
\usepackage{graphicx}
\usepackage{amsmath}
\def\a{\alpha}                         
\def\b{\beta}   
\def\g{\gamma}  
\def\d{\delta}

\def\k{\kappa}

\def\m{\mu}

\def\p{\pi}  
   
\def\s{\sigma}

\def\D{\Delta}

\def\beq{\begin{equation}}         \def\eeq{\end{equation}}
\def\beqa{\begin{eqnarray}}              \def\eeqa{\end{eqnarray}}
\def\be{\begin{eqnarray}}                 \def\ee{\end{eqnarray}}

\def\ba{\begin{array}}        \def\ea{\end{array}}
\def\bi{\begin{itemize}}        \def\ei{\end{itemize}}

\def\nn{\nonumber \\ }

\def\sq#1{\sqrt{#1}}    
\def\exp#1{e^{#1}}                    
\def\float#1(#2){#1\times 10^{#2}}  
\def\err#1(#2){#1 \pm #2}  
 
\def\order#1{\ensuremath{{\cal O}(#1)} }

\def\bar#1{\overline{#1}}  
\def\bra#1{\ensuremath{ \left\langle #1 \right| } }  
\def\ket#1{\ensuremath{ \left| #1 \right\rangle} }  

\def\NPB#1#2#3    { Nucl. Phys. {\bf B#1}, #2 (#3)}
\def\npps#1#2#3   { Nucl. Phys. Proc. Suppl. {\bf #1}, #2 (#3)}
\def\PLB#1#2#3    { Phys. Lett. {\bf B#1}, #2 (#3)}
\def\PRD#1#2#3    { Phys. Rev. {\bf D#1}, #2 (#3)}
\def\prep#1#2#3   { Phys.Rep. {\bf #1} ,#2 (#3)}
\def\PRL#1#2#3    {{ Phys.~Rev.~Lett. }~{\bf #1} ,~#2~(#3)}
\def\ijm#1#2#3    { Int. j. Mod. Phys.{\bf A#1}  ,#2 (#3)}
\def\mpla#1#2#3   { Mod. Phys. Lett. {\bf A#1} ,#2 (#3)}
\def\zpc#1#2#3    { Zeit. f{\"u}r Physik {\bf C#1} ,#2 (#3)}

\def\Npb#1#2#3    { Nucl. Phys. {\bf B#1}, #3 (#2)}
\def\Npps#1#2#3   { Nucl. Phys. Proc. Suppl. {\bf #1}, #3 (#2)}
\def\Plb#1#2#3    { Phys. Lett. {\bf B#1}, #3 (#2)}
\def\Prd#1#2#3    { Phys. Rev. {\bf D#1}, #3 (#2)}
\def\Prep#1#2#3   { Phys. Rep. {\bf #1} ,#3 (#2)}
\def\Prl#1#2#3    { Phys. Rev. Lett. {\bf #1} ,#3 (#2)}
\def\Ijm#1#2#3    { Int. j. Mod. Phys.{\bf A#1}  ,#3 (#2)}
\def\Mpla#1#2#3   { Mod. Phys. Lett. {\bf A#1} ,#3 (#2)}
\def\Zpc#1#2#3    { Zeit. f{\"u}r Physik {\bf C#1} ,#3 (#2)}

\begin{document}
\title{Isospin relation and SU(3) breaking effects of \\ strong phases in Charmless B decays}
\author{Yue-Liang Wu} \email[Email: ]{ylwu@itp.ac.cn}
\affiliation{ Institute of Theoretical Physics, Chinese Academy of
Science, Beijing 100080, China }
\author{Yu-Feng Zhou} \email[Email: ]{zhou@theorie.physik.uni-muenchen.de}
\affiliation{Ludwig-Maximilians-University Munich, \\
Sektion Physik. Theresienstra$\beta$e 37, D-80333. Munich, Germany}
\begin{abstract}
  Isospin and flavor SU(3) relations in charmless hadronic B decays $B\to \pi
  \pi, \pi K$ are investigated in detail with paying attention to the SU(3)
  symmetry breaking effects in both amplitudes and strong phases. In general,
  the isospin and the flavor SU(3) structure of the effective Hamiltonian
  provide several relations among the amplitudes and strong phases. Whereas a
  global fit to the latest data shows that some relation seems not to be
  favorable for a consistent explanation to the current data within the standard
  model (SM). By considering several patterns of SU(3) breaking, the amplitudes
  and the corresponding strong phases are extracted and compared with the
  theoretical estimations.  It is found that in the case of SU(3) limits and
  also the case with SU(3) breaking only in amplitudes, the fitting results lead
  to an unexpected large ratio between two isospin amplitudes
  $a^{c}_{3/2}/a^{u}_{3/2}$, which is about an order of magnitude larger than
  the SM prediction.  The results are found to be insensitive to the weak phase
  $\gamma$.  By including SU(3) breaking effects on the strong phases, one is
  able to obtain a consistent fit to the current data within the SM, which
  implies that the SU(3) breaking effect on strong phases may play an important
  role in understanding the observed charmless hadronic B decay modes $B\to \pi
  \pi$ and $\pi K$. It is possible to test those breaking effects in the near future
 from more precise measurements of direct CP violation in B factories.
\end{abstract}
\pacs{13.25.Hw, 11.30.Er, 12.15.Hh, 12.60.Fr} 
\maketitle
%
%
%
%
%
%
%

\section{Introduction }
B meson physics and CP violation are the central topics of the present day
particle physics. Recently, exciting experimental results are reported from two
B factories at SLAC and KEK. One of the angles $\b$ in the unitarity triangle of
Cabbibo-Kobayashi-Maskawa (CKM) matrix elements is determined through decay mode
$B\to J/\psi K_{S}$ ~\cite{Aubert:2002ic,Abe:2002bx} with a good precision and
found to be consistent with the other indirect measurement within Standard Model
(SM) \cite{Hocker:2001xe}.  The recent preliminary measurements of time
dependent CP violation in other channels such as $B\to \phi K_{S}$
\cite{Aubert:2002nx,Abe:2002bx} also provide us useful information for an
independent determination of the weak phase $\b$ and for probing new physics
beyond the SM.  Besides mixing induced CP violation, rare B decays and direct CP
violations are also of great importance in determining other weak phase angles
of the unitarity triangle and testing the Kobayashi-Maskawa (KM) mechanism
~\cite{kobayashi:1973fv} in SM.  With the successful running of B factories,
higher precision data on the rare hadronic B decay modes such as $B\to \pi\pi,
\pi K$
~\cite{Aubert:2002jm,Aubert:2002jb,Aubert:2002jj,Casey:2002yd,Abe:2001nq,Cronin-Hennessy:2000kg}
have been obtained, which provide us good opportunities to extract the weak
phase angle $\g$, to test the theoretical approaches for evaluating the hadronic
transition matrix elements and to explore new physics beyond the SM.

On the theoretical side, great efforts have been made to improve the
calculations of hadronic matrix elements. The recently proposed methods such as
QCD Factorization ~\cite{Beneke:1999br,Beneke:2000ry} and pQCD approach
~\cite{Keum:2000ph,Keum:2000wi} have been extensively discussed.  From those
methods, useful information of weak phase angles such as $\g$ can be
extracted\cite{Beneke:2001ev,Keum:2002ri}.  Other approaches which are based on
flavor isospin and SU(3) symmetries are still helpful and important
~\cite{zeppenfeld:1981ex,Savage:1989jx,Gronau:1995hm,He:1998rq,Paz:2002ev} .
The advantage of this kind of approaches is obvious that they are model
independent and more convenient in studying the interference between weak and
strong phases.  Recently the flavor isospin and SU(3) symmetries in charmless B
decays are studied by using global fits to the experiment data
~\cite{Zhou:2000hg,He:2000ys,Fu:2002nr}.  In a general isospin decomposition,
there exist a lot of independent free parameters.  By considering the flavor
isospin and SU(3) symmetries, the number of parameters is greatly reduced and
the method of global fit becomes applicable.
Through direct fit, the isospin or SU(3) invariant amplitudes as well as the
corresponding strong phases can be extracted with a reasonable precision . The
early results \cite{Zhou:2000hg} have already indicated some unexpected large
isospin amplitudes and strong phases .  The fitted amplitudes and strong phases
can also provide useful information for the weak phase $\g$ ~\cite{He:2000ys}.
However unlike isospin symmetry, the flavor SU(3) symmetry is known to be broken
down sizably \cite{Gronau:1998fn,Gronau:2000pk}.  The ways of introducing SU(3)
breaking may have significant influence on the final results. In the usual
considerations, the main effects of SU(3) breaking are often taken into
accounted only in the amplitudes.  To be more general, the study of SU(3)
breaking including strong phases is necessary.



In this paper, we begin with the general isospin and flavor SU(3) relations in
charmless hadronic B decays $B\to \pi \pi, \pi K$.  By using a general isospin
decomposition, isospin invariant amplitudes are determined from latest data
through global fit. Different patterns of SU(3) breaking in both amplitudes and
strong phases are studied in detail.
It is observed that in the SU(3) limit the current data suggest a large
violation of a isospin relation which is associated with the electroweak penguin
diagrams in SM.  The results is found to be insensitive to the value of $\gamma$
when its value lies in the range $60^{\circ} \lesssim \gamma \lesssim
120^{\circ}$.  The inclusion of SU(3) breaking effects, especially the one in
the strong phases can improve the agreement between experiment and theory.
%

\section{a general isospin decomposition and isospin relations  }
The isospin symmetry is a good symmetry, it is helpful to start from a pure
isospin discussion and then include flavor SU(3) symmetry and its breaking
effects in the next step .The effective Hamiltonian for $\D S=0$ nonleptonic B
decays is given by
\be\label{EH}
H_{eff}=\frac{G_{F}}{\sq{2}} \sum_{q=u,c} \lambda_{q}
                       \left(
                              C_{1} O^q_{1}+ C_{2} O^q_{2} + \sum_{i=3}^{10}
                              C_{i}O_{i}
                       \right),
\ee
with $\lambda_{q}=V_{qb}V^{*}_{qd}$ is the products of CKM matrix elements and
the operators are
\begin{align}
O^{q}_{1} &=(\bar d_{\a} q_{\a} )_{V-A} (\bar q_{\b} b_{\b})_{V-A},
&O^{q}_{2} &=(\bar d_{\b} q_{\a})_{V-A} (\bar q_{\b} b_{\a})_{V-A},
\nn
O_{3} &=\sum_{q}(\bar q_{\a} q_{\a} )_{V-A} (\bar d_{\b} b_{\b})_{V-A},
&O_{4} &=\sum_{q}(\bar q_{\a} q_{\b} )_{V-A} (\bar d_{\b} b_{\a})_{V-A},
\nn
O_{5} &=\sum_{q}(\bar q_{\a} q_{\a} )_{V+A} (\bar d_{\b} b_{\b})_{V-A},
&O_{6} &=\sum_{q}(\bar q_{\a} q_{\b} )_{V+A} (\bar d_{\b} b_{\a})_{V-A},
\nn
O_{7} &=\frac{3}{2}\sum_{q} e_{q}(\bar q_{\a} q_{\a} )_{V+A} (\bar d_{\b} b_{\b})_{V-A},
&O_{8} &=\frac{3}{2}\sum_{q} e_{q}(\bar q_{\a} q_{\b} )_{V+A} (\bar d_{\b} b_{\a})_{V-A},
\nn
O_{9} &=\frac{3}{2}\sum_{q} e_{q}(\bar q_{\a} q_{\a} )_{V-A} (\bar d_{\b} b_{\b})_{V-A},
&O_{10} &=\frac{3}{2}\sum_{q} e_{q}(\bar q_{\a} q_{\b} )_{V-A} (\bar d_{\b} b_{\a})_{V-A},
\end{align}
where $O^{u(c)}_{1,2}$,  $O_{3,\dots, 6}$  and $O_{7,\dots ,10}$
are related to tree, QCD penguin and  electroweak penguin diagrams respectively.

The final states of $\pi\pi$ have isospin of $2$ and $0$.
 Let us define the isospin amplitudes $A_2$ and $A_0$ as follows
\be
A_{2} &\equiv& \bra{\pi\pi,I=2} H^{3/2}_{eff} \ket{B}
=\lambda_{u} a^{u}_{2}\exp{i\d^{u}_{2}}+\lambda_{c} a^{c}_{2}\exp{i\d^{c}_{2}},
\nn
A_{0} &\equiv& \bra{\pi,\pi,I=0} H^{1/2}_{eff} \ket{B}
=\lambda_{u} a^{u}_{0}\exp{i\d^{u}_{0}}+\lambda_{c} a^{c}_{0}\exp{i\d^{c}_{0}},
\ee
where $a^{q}_{I}, (q=u,c \mbox{ and } I=2, 0)$ are the amplitudes associated
with $\lambda_{q}$.
The QCD penguin operators  $O_{3,\dots , 6}$
have isospin of $\D I=1/2$.  But  the other operators may have
more isospin components. Taking $O^u_{1}=(\bar d u)_{V-A} (\bar u b)_{V-A}$ as
an example, the isospin decomposition gives
$\mathbf{\bar{2}}\otimes \mathbf{2} \otimes \mathbf{2}= \mathbf{4} \oplus \mathbf{2^{'}} \oplus \mathbf{2}$. Thus it contains
a  $\D I=3/2$ and two independent  $\D I=1/2$ isospin invariant operators.
Let us denote them as $\order{3/2}, \order{1/2}$ and ${\mathcal O}'(1/2) $
respectively, then the other operators can be decomposed in the same way, for example:
\be\label{isospinOPE}
O^u_{1}&=&\frac{1}{3}[  \order{3/2}- \order{1/2} +2{\mathcal O}'(1/2) ],
\nn
O^u_{2}&=&\frac{1}{3}[  \order{3/2}+ 2\order{1/2} -{\mathcal O}'(1/2) ],
\nn
O_{3}&=&\order{1/2} \ \ \ \  O_{4}={\mathcal O}'(1/2),
\nn
O_{9}&=&\frac{3}{2}O_{1}-\frac{1}{2}O_{3}
              =\frac{1}{2}[\order{3/2}-2\order{1/2} +2{\mathcal O}'(1/2)],
\nn
O_{10}&=&\frac{3}{2}O_{2}-\frac{1}{2}O_{4}
              =\frac{1}{2}[\order{3/2}+2\order{1/2}-2{\mathcal O}'(1/2)].
\nn
\ee

Among those operators, $\order{3/2}$ has the highest isospin $\D I=3/2$.
In the decays $B\to \pi\pi$ it is the {\em only} operator which can contribute
to the final isospin $I=2$ states.  Rewrite the effective Hamiltonian in terms of
isospin invariant operators in Eq.(\ref{isospinOPE}) and pick up the isospin
2 parts, one finds
\be
A_{2}=\frac{G_{F}}{\sq{2}}
\left[
   \frac{1}{3} \lambda_{u}(C_1+C_2+C_9+C_{10})) + \frac{1}{2} \lambda_{c}(C_9+C_{10}))
\right] \bra{I=2} \order{3/2}\ket{B},
\ee
and
\be\label{relation}
\frac{a^c_2}{a^u_2} \equiv R_{EW}=\frac{3}{2}\cdot \frac{C_9+C_{10}}{C_1+C_2+C_9+C_{10}}.
\ee
Taking the Wilson coefficients at $\m=m_{b}$, one has $C_1=1.144, C_2=-0.308,
C_9=-1.28 \a, C_{10}=0.328 \a$.  Thus \be R_{EW}=-1.25\times 10^{-2}, \ \ 
\mbox{and} \ \ \d^{c}_{2}=\d^{u}_{2}.  \ee

This relation is well known and has been extensively discussed
\cite{Neubert:1998pt,He:1998rq,He:2000ys,Fu:2002nr,Paz:2002ev}.  Here we would
like to emphasize the importance of this relation in a model independent
analysis, namely:

1) The relation is obtained without the knowledge of the matrix element
$\bra{I=2} \order{3/2}\ket{B}$. It only depends on the isospin structure of the
effective Hamiltonian and the final states.  Thus it is independent of any model
calculations, such as naive factorization or pQCD factorization etc.

2) It can not be affected by the final state inelastic rescattering processes
with lower isospin as it is only related to the highest isospin component $\D
I=3/2$.  For example, it is expected that the processes of $B\to DD \to \pi\pi$
may be considerable in B decays\cite{Kamal:1999rn,Xing:2000pc}. Whereas the
effective Hamiltonian of $B\to DD$ have isospin $1/2$, its contribution to final
state with $I=2$ vanishes, thus the above mentioned relation remains unchanged.
The elastic rescattering process $B\to \pi\pi \to \pi\pi$ can contribute to the
highest isospin amplitude, but their effects can be absorbed into the effective
value of $\bra{I=2} \order{3/2}\ket{B}$ and will not affect the value of
$R_{EW}$ which is the ratio of two isospin amplitudes sharing the same matrix
elements.  Thus this relation is less likely to be modified in the presence of
final state interaction (FSI).

3) In the usual digram language, the decay $B\to \pi^-\pi^0$ receives
contributions from several diagrams, i.e.,
$A(\pi^-\pi^0)=T+T^{C}+P_{EW}+P^C_{EW}$ (here "$T$" and "$P_{EW}$" stand for
tree and electroweak penguin diagrams, the superscription ``$C$'' stands for the
corresponding color suppressed one). It is expected that the interference
between them may result in a small direct CP violation.  However from relation
$\d^{c}_{2}=\d^{u}_{2}$, it is easy to see that as long as the isospin symmetry
is imposed, there is $no$ direct CP violation in $B\to \pi^-\pi^0$.  This
conclusion purely relies on the isospin considerations and thus looks quite
robust. A similar observation was also made within SU(3) symmetry in
Ref.\cite{Fu:2002nr}.  However when comparing to the possible nonnegligible
SU(3) breaking effects, the conclusion based on isospin symmetry seems more
reliable.

4) The value of $R_{EW}$ is the ratio between the electroweak penguin and tree
diagrams. It is then sensitive to new physics effects beyond the SM in
electroweak penguin sector. The new physics effects on $R_{EW}$ have been
discussed in Refs.\cite{He:1999az,Grossman:1999av,Ghosh:2002jp,Xiao:2002mr}, it
seems quite sensitive to several new physics models.  A precise determination of
$R_{EW}$ from experiments may be helpful to single out possible new physics or
study flavor symmetry breaking in charmless B decays.  To describe the
possibility that the value of $R_{EW}$ extracted from experiments could be
different from the SM calculations, we introduce a factor $\k$ as follows 
\be
R^{exp}_{EW}=\k \cdot R_{EW} \simeq -0.0125\cdot \k,
\ee
 where $R^{exp}_{EW}$
stands for its value extracted from experiments and obviously $\k=1$ in SM.


Let us consider the operators with lower isospins.  Note that the operators
$O^c_{1}$ and $O^c_{2}$ have isospin of $1/2 $. As final states $\pi\pi$ are
charmless and have isospin $2$ and $0$, those operators can not contribute
directly.  However, through inelastic final state interaction (FSI) processes
such as $B\to DD \to \pi\pi$, their contributions to the final state with
isospin $0$ may be non-negligible. At present stage, there is no good
theoretical estimation of such kind of processes.
The operator $O_{5}$ and $O_{6}$ also have isospin $1/2$ but with different
Lorenz structure. In general, the matrix elements of $O_{5,6}$ are different
from $O_{3,4}$. Thus the isospin amplitude $A_0$ receives contributions from
many different operators with the same isospin $1/2$.  The matrix elements of
those operators may develop different strong phases.  Although for each operator
there exist relations between $\lambda_{u}$ and $\lambda_{c}$ parts, there is no
simple relation for their sum. In the most general case $a^{u}_{0}$ and
$a^{c}_{0}$ are independent of each other and $\d^{u}_{0} \neq \d^{c}_{0}$.

A similar discussion can be made in  decay modes $B\to \pi K$, where
the effective Hamiltonian has isospin $\D I=1,0$.
In this case one can define three isospin components
\be
A_{3/2} &\equiv& \bra{\pi K,I=3/2} H^{\D I=1}_{eff} \ket{\bar B^0}
=\lambda_{u} a^{u}_{3/2}\exp{i\d^{u}_{3/2}}+\lambda_{c} a^{c}_{3/2}\exp{i\d^{c}_{3/2}},
\nn
A_{1/2} &\equiv& \bra{\pi K,I=1/2} H^{\D I=0}_{eff} \ket{\bar B^0}
=\lambda_{u} a^{u}_{1/2}\exp{i\d^{u}_{1/2}}+\lambda_{c} a^{c}_{1/2}\exp{i\d^{c}_{1/2}},
\nn
B_{1/2} &\equiv& \bra{\pi K,I=1/2} H^{\D I=1}_{eff} \ket{B^-}
=\lambda_{u} b^{u}_{1/2}\exp{i\d'^{u}_{1/2}}+\lambda_{c} b^{c}_{1/2}\exp{i\d'^{c}_{1/2}}
\ee
As there are two kind of Lorenz structure $(\bar q q)(\bar s b)$ and
$(\bar s q)(\bar q b)$ with the same isospin, there are two independent
operators with highest isospin $\D I=1$. One can not construct a similar
relation of Eq.(\ref{relation}) within isospin symmetry. However, as
it will be discussed below, one can obtain from SU(3) symmetry some useful relations.

%

From the above discussions the general form of isospin decomposition of the
decay amplitudes for $B\to \pi\pi (\pi K)$ decays reads 
\be A^{\pi\pi(\pi K)}=
\lambda^{(s)}_{u} A^{\pi\pi(\pi K)}_u+\lambda^{(s)}_c A^{\pi\pi(\pi K)}_c, 
\ee
where $\lambda^{(s)}_{u}=V_{ub}V^*_{ud(s)}$,
$\lambda^{(s)}_{c}=V_{cb}V^*_{cd(s)}$ and
\be\label{isospinPP}
A_{q}^{\pi^-\pi^+} &=&\sqrt{\frac{2}{3}} \ampq{a}{0}+\sqrt{\frac{1}{3}}
\ampq{a}{2},
\nn 
A_{q}^{\pi^0\pi^0} &=&\sqrt{\frac{1}{3}} \ampq{a}{0} -
\sqrt{\frac{2}{3}}\ampq{a}{2},
\nn 
A_{q}^{\pi^-\pi^0}&=&-\sqrt{\frac{3}{2}},
\ampq{a}{2},
\ee 
and 
\be\label{isospinPK} A_{q}^{\pi^+K^-} &=&
\sqrt{\frac{2}{3}}\ampq{a}{1/2} + \sqrt{\frac{1}{3}}\ampq{a}{3/2},
\nn
A_{q}^{\pi^0\bar{K^0}}&=& \sqrt{\frac{1}{3}}\ampq{a}{1/2} - \sqrt{\frac{2}{3}}
\ampq{a}{3/2} ,
\nn 
A_{q}^{\pi^0K^-}&=&-\sqrt{\frac{1}{3}}\ampq{b}{1/2} -
\sqrt{\frac{2}{3}} \ampq{a}{3/2},
\nn
A_{q}^{\pi^-\bar{K^0}}&=&\sqrt{\frac{2}{3}}\ampq{b}{1/2} - \sqrt{\frac{1}{3}}
\ampq{a}{3/2}.
\ee 
with $q=u,c$. By using relation Eq.(\ref{relation}) and
dropping a global phase which is unphysical, there are totally 17 free
parameters.

\section{flavor SU(3) symmetry and its breaking effects}
The advantage of the isospin decomposition allows one to study $SU(3)$ relations
and $SU(3)$ breaking effects in a convenient way that the isospin symmetry
clearly persists. In SU(3) limit with annihilation topology ignored, the isospin
amplitudes satisfy the following relations: \be\label{SU3relation}
a^{u}_{0}\exp{i\d^{u}_{0}}&=& a^{u}_{1/2}\exp{i\d^{u}_{1/2}}, \nn
a^{c}_{0}\exp{i\d^{c}_{0}}&=& a^{c}_{1/2}\exp{i\d^{c}_{1/2}}, \nn
a^{u}_{2}\exp{i\d^{u}_{2}}&=& a^{u}_{3/2}\exp{i\d^{u}_{3/2}}, \nn
a^{c}_{2}\exp{i\d^{c}_{2}}&=& a^{c}_{3/2}\exp{i\d^{c}_{3/2}}. \ee If these
relations are adopted, the number of free parameters is reduced to be nine.
From Eq. (\ref{relation}) and the above relation, one finds that
\be\label{relation2}
\frac{a^{c}_{3/2}}{a^{u}_{3/2}}=\frac{a^{c}_{2}}{a^{u}_{2}}=R_{EW}.  \ee Thus
the highest isospin amplitudes for the $B\to \pi K$ decays satisfy the same
relation as the one in the $B\to \pi\pi$ decay.
When SU(3) breaking effects are considered, the above relations have to be
modified. At present stage, it is not very clear how to describe the SU(3)
breaking effects. a widely used approach is introducing a breaking factor $\xi$
which characterizes the ratio between $B\to \pi K$ and $\pi\pi$ decay
amplitudes, i.e., 
\be\label{simpleSU3brk} 
a^{u(c)}_{1/2} = \xi a^{u(c)}_{0} , \ 
\ \ a^{u(c)}_{3/2} = \xi a^{u(c)}_{2},
\ee 
but their strong phases are assumed to
remain satisfying the SU(3) relations 
\be\label{EqualPhase} 
\delta^{u(c)}_{1/2}
= \delta^{u(c)}_{0} , \ \ \ 
\delta^{u(c)}_{3/2} = \delta^{u(c)}_{2}.
\ee
Typically $\xi=f_K/f_\pi\simeq 1.23$ with $f_\pi$ and $f_K$ being the pion and
kaon meson decay constants, which comes from the naive factorization
calculations.  It is easy to see that this pattern of SU(3) breaking is a quite
special one.  The value of $\xi$ is highly model dependent. It can only serve as
an order of magnitude estimation and it is even not clear whether a single
factor can be applied to all the isospin amplitudes. The equal strong phase
assumption implies that the SU(3) breaking effects on strong phase are all
ignored, which may be far away from the reality.  In a more general case, all
the strong phases could be different when SU(3) is broken down. The breaking
effects on strong phases may have significant effects on the prediction for the
direct CP violations in those decay modes.

To describe the possible
violations of relations in eq. (\ref{EqualPhase}) or the SU(3) breaking effects
on strong phases, we may introduce the following phase
differences $\D^{q}_{I} (q=u,c$ and $I=3/2, 1/2 )$:
\be
\delta^{q}_{0}=\delta^{q}_{1/2}+\D^{q}_{1/2}, \ \ \
\delta^{q}_{2}= \delta^{q}_{3/2} + \D^{q}_{3/2} \ \ \  \ \ (q=u,c).
\ee
On the other hand, the SU(3) breaking effects in amplitudes may also be given in a more
general way
 \be\label{nosimpleSU3brk}
 a^{q}_{1/2} = \xi^q a^{u(c)}_{0} , \ \ \ a^{q}_{3/2} = \xi^q    a^{q}_{2} \ \ \  \ \ (q=u,c)
\ee
The SU(3) limit corresponds to the case that all $\D^{q}_{I}$ vanish and $\xi^q =1$.
In general, the simple SU(3) breaking pattern in Eq.(\ref{simpleSU3brk}) and
Eq.(\ref{EqualPhase}) may become unreliable.
Note that in the
simple SU(3) breaking pattern in Eq.(\ref{simpleSU3brk}) and
Eq.(\ref{EqualPhase})   the relation of Eq.(\ref{relation2})
remains to be unchanged as it is the ratio of two isospin amplitudes.
The calculation based on the naive factorization shows a very small breaking of
this relation \cite{Neubert:1998pt}. For simplicity,
in the following discussions we should not discuss the
violation of amplitude relation in Eq.(\ref{relation2}), but the
exact value of $R_{EW}$ (i.e. $R^{exp}_{EW}$ or $\k$) will
be studied in detail and also the possible violation of strong phases will be discussed.

Without any model calculations, all the isospin amplitudes and the strong phases
are unknown free parameters. Those parameters can in principle be extracted from
the experimental data, namely through a global fit of the data on branching
ratios as well as direct CP violations of the related decay modes. The precision
of the fitted parameters depend on the precision of the current data. Especially
for the values of strong phases which strongly depend on the measurements of
direct CP violation.
\section{value of  {\large $\mathbf{\k}$} in different patterns of SU(3) breaking}
The basic idea of the global fit is the maximal likelihood or minimal
$\chi^2$ method.  For a set of  measurements on
observables $Y_{i} ( i=1,m) $ which contain $n$ parameters
$\a_{j} (j=1,n)$, a quantity $\chi^2$ is constructed as follows
\be
\chi^2=\sum_{i} \left( \frac{Y_{i}^{th}(\a_{j})-Y_{i}^{ex}}{\s_{i}}
\right)^2,
\ee
where $Y_{i}^{th}(\a_{j})$ and $Y_{i}^{ex}$ are corresponding to the theoretical
and experimental values of the observable $Y_{i}$ which, in our present case, is
a decay rate or direct CP violation in charmless B decays.  $\s_{i}$ is the
corresponding error of the measurements.  The set of $\a_{j}$ which minimize the
value of $\chi^2$ corresponds to the best estimated value for $\a_{j}$.

From the general isospin decomposition of Eqs.(\ref{isospinPP}) and (\ref{isospinPK})
and the isospin relation of Eq.(\ref{relation}) as well as the
SU(3) relation (\ref{SU3relation}), there are nine free parameters left
$$
a^{u}_{1/2},\  \d^{u}_{1/2},\  a^{u}_{3/2},\  \d^{u}_{3/2},\ a^{c}_{1/2},\
b^{u}_{1/2},\  \d'^{u}_{1/2},\  b^{c}_{1/2},\ \d'^{c}_{1/2}
$$
Here we set $\d^c_{1/2}=0$ as a phase convention since
one of the phases can always be removed without affecting the physics.
All the other phases are defined within the range $(-\pi, +\pi)$.
The theoretical values of those parameters have been calculated in
Ref.\cite{Zhou:2000hg} which are normalized to the branching ratio of B decays
and in units of $10^{-3}$. The calculation shows a hierarchical structure with
$a^{u}_{I} \gg a^{c}_{I}$ which corresponds to $T \gg P$ in diagram language.
The value of $b^{u}_{1/2}$ is found to be significantly smaller than
$a^{u}_{1/2}$. Due to further suppression of small CKM matrix element, the
contribution of $b^{u}_{1/2}$ is quite small. Unlike $a^{u}_{1/2}$ which is
connected to $B\to \pi\pi$ amplitudes through SU(3) symmetry, $b^{u}_{1/2}$ only
appears in the charged decay modes $B\to \pi^0 K^-, \pi^- \bar{K}^0$, its value
only has a little effect on the fit of other parameters. It have been checked
that the fitted values for other parameters are quite stable even under the
significant changes of $b^{u}_{1/2}$\cite{Zhou:2000hg}. Thus it is a good
approximation to fix $b^{u}_{1/2}$ at its theoretical value $b^{u}_{1/2}\simeq
416$ and $ \d'^{u}_{1/2} \simeq 0$.  With this approximation, only seven free
parameters are left in the flavor SU(3) symmetry limit.

In the following section, the global fit of charmless B decay modes are made
under several different cases of SU(3) breaking. The latest data of the decays
$B\to \pi\pi, \pi K$  used in the fits are summarized  in table \ref{br}.
Among other parameters concerning CKM matrix elements, the most uncertain one is
the weak phase $\gamma$. The most recent updated global fit on CKM matrix
elements is summarized in Ref.\cite{Parodi:ICHEP02}, which gives
$\bar\rho=0.199\pm0.04$ and $\bar\eta=0.345\pm0.026$, corresponds to $\g\simeq
60^\circ$.  In this work, the various SU(3) relations are examined with the
value of $\gamma$ varying from $60^{\circ}$ to $120^{\circ}$.
%
%
For a concrete illustration, three interesting cases are discussed:\\
\noindent{\bf Case 1}.
The value of $\gamma$ is taken to be $60^{\circ}$ and $\k$ is fixed to be unity.
The global fit is done with $\xi=1$ and $\xi=f_{K}/f_{\pi}=1.23$ which
corresponds to the exact SU(3) symmetry and the simple SU(3) breaking. The
results are shown in the first (a) and second (b) column of table
\ref{simpleFit}.  In both cases large strong phases are resulted with the
minimal of $\chi^2$ around 5.8(9.2) for $\xi=1(1.23)$.  From the fit result, the
corresponding direct CP violation can also be obtained.  The best fitted direct
CP violation for example, in case (c) is given by
\begin{align}\label{CP1}
A_{CP}(\pi^{+}\pi^{-}) & \simeq0.3
&A_{CP}(\pi^{0}\pi^{0}) &\simeq 0.4
\nn
A_{CP}(\pi^{+} K^-) &\simeq -0.1
&A_{CP}(\pi^{0}\bar K^{0}) &\simeq -0.1
\nn
A_{CP}(\pi^{0} K^-) & \simeq -0.0
&A_{CP}(\pi^{-} \bar K^0) &\simeq 0.1
\end{align}
%

\begin{table}[htb]
\caption{gloal fit of isospin amplitudes and strong phases in charmless B decays with 
$\gamma=60^\circ$}
\begin{center}
\begin{ruledtabular}
\begin{tabular}{ccccc}
parameter & value(a) &value(b)&value(c)&value(d)\\\hline
 $a^u_{1/2}$&$     517.0^{+        81.5}_{       -80.6}$&$401.5^{+       125.1}_{      -205.2}$&$       293.8^{+        58.8}_{       -55.9}$&$       415.0^{+        77.8}_{       -77.8}$\\          
 $\d^u_{1/2}$&$     2.42^{+         0.3}_{        -0.2}$&$    1.22^{+         0.3}_{        -1.5}$&$         0.7^{+         0.5}_{        -0.3}$&$         0.6^{+         0.4}_{        -0.3}$\\ 
 $a^c_{1/2}$&$     0.85^{+         2.9}_{        -2.9}$&$      -0.28^{+         2.9}_{        -2.8}$&$      -2.62^{+        2.46}_{       -1.97}$&$      1.18^{+        1.24}_{       -0.36}$\\
 $a^u_{3/2}$&$     536.8^{+        38.6}_{       -41.8}$&$       667.2^{+        48.1}_{       -51.8}$&$       432.4^{+        48.8}_{       -51.7}$&$       545.9^{+        51.4}_{       -54.6}$\\ 
 $\d^u_{3/2}$&$     3.09^{+         0.3}_{        -0.3}$&$      0.01^{+         1.2}_{        -0.3}$&$       1.43^{+         0.1}_{        -0.1}$&$       1.43^{+         0.1}_{        -0.1}$\\ 
 $b^c_{1/2}$&$    -141.0^{+         4.2}_{        -4.2}$&$      -148.0^{+         4.2}_{        -4.1}$&$      -132.1^{+        15.5}_{       -10.9}$&$      -127.3^{+        16.6}_{       -12.1}$\\
 $\d'^u_{1/2}$&$       2.8^{+         0.4}_{        -0.5}$&$       -0.28^{+         1.0}_{        -0.4}$&$        -0.1^{+         0.2}_{        -0.2}$&$        -0.2^{+         0.2}_{        -0.2}$\\
$\xi$ & 1.0(fix) & 1.23(fix) & 1.0(fix) &1.23(fix)\\
$\k$  & 1.0(fix) & 1.0 (fix) &  $12.0^{+         5.3}_{        -4.4}$&$10.7^{+         3.6}_{        -3.2}$\\\hline
$\chi^{2}_{min}$& 5.8 & 9.2 & 0.61 &0.85 
\end{tabular}
\end{ruledtabular}
\label{simpleFit}
\end{center}
\end{table}
 
\vskip 12pt
\noindent{\bf Case 2}\\
{\bf a)}
The value of $\gamma$ is fixed at $60^{\circ}$ but the value of $\k$ taken as a free parameter which
is to be determined from global fit with $\xi=1.0$ and 1.23.  The results are shown in the third (c)
and fourth (d) column of table \ref{simpleFit}.  In this case, the best fitted value of $\k$ is found
to be quite large with  a very low $\chi^{2}_{min}\simeq 1$, which indicates that a large  $\k$  
is in a better agreement with the  current  data.   The numerical results for the best
fitted value are
\be
\k=12.0 (10.7) \ \ \ \ \ \mbox{for } \xi=1.0 (1.23),
\ee
which  is about an order of magnitude larger than the expected one from the SM.  While the results
confirm our earlier numerical results obtained in Ref.\cite{Zhou:2000hg} where the equal phase
assumption such as $\d^{u}_{0}=\d^{c}_{0}$ has been adopted to reduce the number of free parameters.
Here the fit is made in the most general case where $\d^{u}_{i} \neq\d^{c}_{i}$ and thus more
reliable.

{\bf b)} To examine whether the above results hold only for a particular value of
the weak phase $\gamma=60^{\circ}$, similar fits are made with $\gamma=75^{\circ},
90^{\circ}, 105^{\circ}$ and $120^{\circ}$.  The results  listed in
Table.\ref{gammas}  clearly show that the $\gamma$ dependence is rather weak.  For all
the values of $\gamma$ the best fitted values of $\k$ are found to be large.
Even at $\gamma=120^{\circ}$ the best fitted value of $\k\simeq 7.4$ is still
much higher than unity.
While the global fit based on naive factorization and QCD factorization
calculations prefer a large value of $\gamma>
90^{\circ}$\cite{He:1999mn,Beneke:2001ev}, the model independent estimations
show a less sensitivity of weak phase $\g$
\cite{Wu:2000rb,He:2000ys,Zhou:2000hg}.  For example, in our earlier analysises
based on diagram decomposition\cite{Wu:2000rb} and SU(3)
symmetry\cite{He:2000ys} in $B\to \pi\pi, \pi K$ decays, two allowed ranges of
$\gamma$ are found, the one with $\gamma<90^{\circ}$ and the other one with
$\gamma>90^{\circ}$. Both values of $\gamma$ with appropriate strong phases can
reproduce the experimental data.  The resulted large $\k$ which is insensitive
to the weak phase $\gamma$ implies that the breaking effects of flavor SU(3)
symmetry may be considerable.

Let us discuss the possibility of a large $\k$ or $a^{c}_{2(3/2)}$ from the phenomenological point
of view.  It is well known that due to the suppression of small CKM matrix element $V_{ub}$, the
decays $B\to \pi K$ are dominated by QCD penguin diagrams. The naive factorization calculations
indicate that the dominant terms in the decay amplitudes are those with the CKM factor
$\lambda^{s}_{c}$. If $a^{c}_{2(3/2)}$ is negligible small, one finds that $ Br(\pp\km) \simeq
2\cdot Br(\pz\kzbar) $. 
When $a^{c}_{3/2}$ is large, namely $\k$ is large, the interference between
$a^{c}_{1/2}$ and $a^{c}_{3/2}$ will be important.  From Eq.(\ref{isospinPK}) it
follows that when both the amplitude $a^{c}_{3/2}$ and the strong phase
$\d^c_{3/2}-\d^c_{1/2}$ become large, such an interference will enhance the
branching ratios of $B\to \pz\kzbar$ , and suppress the ones of $B\to \pp\km$.
Similarity occurs in the decay mode $B\to \pi^0\pi^0$. As the tree diagram
contributions in this decay mode are color suppressed, the penguin contributions
are more important than the ones in other modes $B\to \pi^-\pi^0$ and
$\pi^-\pi^+$.  Thus large value of $\k$ and $\d^c_{3/2}-\d^c_{1/2}$ will also
enhance the branching ratio of $B\to \pi^0\pi^0$. From the relation of
Eq.(\ref{relation2}) and the definition of $\d^{c}_{1/2}=0$, one has
$\d^c_{3/2}-\d^c_{1/2}= \d^c_{3/2}=\d^{u}_{3/2}$.  As the fitting results also
give large $\d^{u}_{3/2}=1.43(\simeq  80^{\circ})$ for $\xi=1.0$, such an anomaly
is closely related to the observed enhancement of $B\to\pz\kzbar$. For decay
mode $B\to \pi^0\pi^0$, the current data can only give an upper bound of
$Br(B\to \pi^0\pi^{0})<\sci{3.6}{-6}$ \cite{Bona:2003Paris,Unno:2003yx,Aubert:2003qj}, 
however, the primitive measurements from CLEO, Babar and Belle also show an indication of  a large averaged value of $Br(B\to
\pi^0\pi^{0})=\sci{1.96}{-6}$ ( see table \ref{br}), which need to be confirmed
by the future experiments.  From the most recent experimental data in table 
\ref{br}, one has
\be \frac{2
  Br(\pz\kzbar)}{Br(\pp\km)}\simeq1.09,\quad 
\frac{Br(\pz\pz)}{Br(\pi^{-}\pp)}<0.87( \simeq 0.43), 
\ee
which may be compared with the recent theoretical estimations by using QCD
factorization \cite{Beneke:2001ev,Du:2000ff} 
\be \frac{2
  Br(\pi^{0}\kzbar)}{Br(\pp\km)}\simeq0.52,\quad
\frac{Br(\pi^{0}\pi^{0})}{Br(\pi^{-}\p^{+})}\simeq0.01.  
\ee 
It is clear that the data present the unexpected large  ratio for the
decay modes $B\to\pz\kzbar$ , which significantly deviates from the 
QCD factorization predictions. The current data also imply  the probability that  
$B\to \pi^0\pi^0$ could be much larger than the expected one from QCD factorization.

The value of $\k$ is sensitive to the contributions from electroweak penguin
diagrams.  Since many new physics models can give significant corrections to
this sector, it may be helpful to study new physics effects on $\k$.  However,
to explore any new physics effects and arrive at a definitive conclusion for the
existence of new physics from the hadronic decays, it is necessary to check all
the theoretical assumptions and make the most general considerations.  It is
noted that the above results are obtained by assuming SU(3) symmetry with its
breaking only in amplitudes. Therefore, we shall first extend the above results
to a more general case of SU(3) symmetry breaking before claiming any possible
new physics signals.

\vskip 12pt
\noindent{\bf Case 3}\\
{\bf a)} The value of gamma is fixed at $60^{\circ}$ and
the SU(3) breaking effects on
strong phases are turned on, i.e., $\D^{q}_{I}\neq 0$. In this case, it is difficult to extract
those breaking factors with a reasonable precision as we have no enough data
( especially data of direct CP violations ) at hand.  For illustrations,
we then take some typical values for $\D^{q}_{I}$ to show how the best fitted values of $\k$
depend on the ways of SU(3) breaking in strong phases. For simplicity and also to see how
the SU(3) symmetry breaking of each strong phase affects the best fitting value of $\k$,
we take four typical values for each  $\D^{q}_{I}$, i.e., 
$\D^{q}_{I}=-\pi/3, -\pi/6,  +\pi/6$ and  $+\pi/3$, with  others angles being fixed to be zero.
 The numerical results can be seen in table \ref{g60}.
It follows from the table that the inclusion of nonzero $\D^{q}_{I}$
can greatly modify the best fitted value of $\k$. In
some cases, the best fitted values are found to be close to unity.
For example, in cases of $\D^{u}_{1/2}=+\pi/6$, $\D^{c}_{1/2}=+\pi/6, +\pi/3$
and $\D^{c}_{3/2}= +\pi/6, +\pi/3$, the best fitted values of $\k$ are around
1.5 with the minimal $\chi^2_{min}\leq 4$.  The direct CP violation for
$\D^{u}_{1/2}=+\pi/6(\D^{c}_{1/2}= +\pi/6)$ is as follows.
\begin{align}
A_{CP}(\pi^{+}\pi^{-}) &\simeq 0.1(  0.5 ),&
A_{CP}(\pi^{0}\pi^{0}) &\simeq  0.5(0.2 ),
\nn
A_{CP}(\pi^{+} K^-) &\simeq -0.1(-0.1),&
A_{CP}(\pi^{0}\bar K^{0}) &\simeq  -0.2(-0.1 ),
\nn
A_{CP}(\pi^{0} K^-) &\simeq -0.1(-0.0),&
A_{CP}(\pi^{-} \bar K^0) &\simeq  0.1(0.1).
\end{align}
Compared with the ones with SU(3) syemmetry in Eq.(\ref{CP1}), the predicted
values of direct CP violation can be  quite different.
 The above results indicate that if we want to explain the current data within the scope of SM, the
SU(3) breaking effects on strong phases may play an important role.  At present, the calculation of
SU(3) breaking on strong phases is not reliable without well considering the nonperturbative effects
and it is hard to estimate how large it could be. The phenomenological approach adopt in this paper
may provide us some clues to understand the SU(3) symmetry breaking due to final state interactions.

{\bf b)} To illustrate the possible $\gamma$ dependence, two other fits are made
with $\gamma=90^{\circ}$ and $120^{\circ}$. The numerical results are shown in
Table. \ref{g90} and \ref{g120}.  In the case of $\gamma=90^{\circ}$, some
results are found with $\k\simeq 1.0$ and small $\chi^{2}_{min}$.  For example,
for $\D^{u}_{3/2}=-\pi/3,-\pi/6$ and $\D^{c}_{3/2}= -\pi/3, -\pi/6$ the best
fitted value of $\k$ are around 1.0 with the minimal $\chi^2_{min}\simeq 3.0$.
However, in the case of $\gamma=120^{\circ}$, no solution is found with both
small $\chi^{2}_{min}$ and $\k \approx 1$.  As only several typical values of
$\D^{u(c)}_{1/2(3/2)}$ are used in the fit, one should not draw a conclusion
that the case of $\gamma=120^{\circ}$ is not likely to be consistent with
current data even the SU(3) breaking effects of strong phases are taken into
account. However, it is clear that for very large value of $\gamma$, the allowed
parameter space for the strong phase differences $\D^{u(c)}_{1/2(3/2)}$ is much
smaller. 

The results summaried in table \ref{g60}, \ref{g90} and \ref{g120} also indicate
the prefered values of some strong phases. For instance, the best fited value
of $\k$ is found to be insensitive to the value of $\D^{u}_{3/2}$. When $\gamma$
is taken to be $60^{\circ}$ and $90^{\circ}$ the best fitted value of $\k$ is
close to unity for all the values of $\D^{u}_{3/2}$, with a small
$\chi^{2}_{min}$. But for $\D^{u}_{3/2}=-60^{\circ}$, the $\chi^{2}_{min}$ has a
minimal of 2.3(2.1) for $\gamma=60^{\circ}(90^\circ)$. It implies that the
favoured value for $\D^{u}_{3/2}$ should be close to $-60^{\circ}$ from the
current data. Simiarlily, the fit results favour a large negative value of $\D^{c}_{3/2}$
and a small positive $\D^{c}_{1/2}$ and $\D^{u}_{1/2}$.

\section{conclusions}
In summary, we have investigated the isospin and flavor SU(3) relations and their validity in the
charmless hadronic B decays $B\to \pi \pi, \pi K$.
Through a global fit to the latest data, the amplitudes as well as the corresponding strong phases
are extracted with different patterns of SU(3) breaking.
%
%

It has been shown that in the case of SU(3) limits and the case with SU(3) breaking only in
amplitudes, the fitting results require a large value for the ratio of two isospin amplitudes
$a^{c}_{3/2}/a^{u}_{3/2}$. The rescaled ratio $\k$ which is equal to 1 in SM is found in this case
to be
$$
\k=12.0(10.7)   \ \ \ \ \  \mbox{for} \ \ \xi=1.0(1.23) ,
$$
with a minimal  $\chi^2$ around 1. Such a value of $\k$  is  about an order of
magnitude greater than the SM prediction. This results is insensitive to the
weak phase $\gamma$.
The SU(3) breaking effects on strong phases have been studied in several cases.
It has been seen that the best fitted value of $\k$ can significantly be lowed
or even close to the SM prediction $\k = 1$ with
a minimum $\chi^2$ at about 4.
It implies that to understand the current data within SM, the SU(3) breaking effects of the strong
phases must be considered and it is likely to play an important role. The direct test on the SU(3)
breaking of the strong phases require more precise measurements of direct CP violation.  With the
accumulating of the data in B factories, this may become possible in the near future.
\begin{acknowledgments}
This work is supported in part by the Chinese Academy of Sciences and NSFC under Grant
$\#$ 19625514. YFZ acknowledges the  support by Alexander von Humboldt Foundation.
YFZ also thanks G. Buchalla for reading the manuscript and  Z.Z. Xing for
helpful discussions. YLW is grateful to L. Wolfenstein
for very useful discussions concerning the importance of strong phases.
\end{acknowledgments}

\bibliography{reflist}
\bibliographystyle{apsrev}
 
\newpage
\begin{table}[htb] 
\caption{The branching ratios for $B\to PP$ in units of $10^{-6}$ \cite{Bona:2003Paris,Unno:2003yx,Aubert:2003qj,Cronin-Hennessy:2000kg}.}
\begin{ruledtabular}\label{br}
\begin{tabular}{lllll}
 Br and Acp  & CLEO &Belle & Babar    &Averaged \\ \hline
 $Br(  \pi^+\pi^-)$         &    $4.5^{+1.4+0.5}_{-1.2-0.4}$         &     $4.4\pm 0.6\pm 0.3$               &    $4.7 \pm0.6 \pm0.2$                      &            $4.6\pm0.4$\\
 $Br( \pi^0\pi^0)$          &    $<4.4(2.2^{+1.7+0.7}_{-1.3-0.7})$   &   $<4.4(2.9 \pm 1.5 \pm 0.6)$         &    $<3.6(1.6^{+0.7+0.6}_{-0.6-0.3})$        &         $<3.6(1.96\pm0.73)$ \\
 $Br(  \pi^-\pi^0)$         &    $4.6^{+1.8+0.6}_{-1.6-0.7}$         &    $5.3\pm 1.3\pm 1.5$                &    $5.5^{+1.0}_{-0.9}\pm0.6$                &         $5.3\pm0.8$\\
 $Br(  \pi^+K^-)$           &    $18.0^{+2.3+1.2}_{-2.1-0.9}$        &    $18.5\pm1.0\pm0.7$                 &    $17.9\pm 0.9 \pm0.7$                     &        $18.2 \pm 0.8$\\
 $Br(  \pi^0\bar K^0)$      &    $12.8^{+4.0+1.7}_{-3.3-1.4}$        &    $12.6 \pm 2.4\pm 1.4$              &     $10.4^{+1.5}_{-1.5}\pm0.8$               &        $11.5\pm 1.7$\\
 $Br(  \pi^-\bar K^0 )$     &    $18.8^{+3.7+2.1}_{-3.3-1.8}$        &   $22.0\pm1.9\pm1.1$                  &   $17.5^{+1.8}_{-1.7}\pm 1.3$               &        $20.6\pm 1.4$\\
 $Br(  \pi^0K^-)$           &    $12.9^{+2.4+1.2}_{-2.2-1.1}$        &    $12.8 \pm 1.4^{+1.4}_{-1.0}$       &    $12.8^{+1.2}_{-1.1}\pm 1.0$              &       $12.8\pm 1.1$\\
 $A_{CP}(  \pi^-\pi^0)$     &                                        &   $0.31\pm0.31\pm0.05$                &   $-0.03^{+0.27}_{-0.26}\pm 0.10$           &   $0.13\pm0.21$\\
 $A_{CP}(  \pi^+\pi^-)$     &                                        &   $0.94^{+0.25}_{-0.31}\pm0.09$       &   $-0.02\pm0.29\pm0.07$                     &     $0.42\pm0.22$ \\
 $A_{CP}(  \pi^-\bar K ^0)$ &         $  0.18\pm0.24$                &        $0.46\pm0.15\pm0.02$           &        $-0.17\pm0.10\pm0.02$                &      $0.04\pm0.08$\\
 $A_{CP}(  \pi^0K^-)$       &         $-0.29\pm0.23$                 &        $-0.04\pm0.19\pm0.03$          &       $-0.09\pm0.09\pm0.01$                  &       $-0.1\pm0.07$\\
 $A_{CP}(  \pi^+K^-)$       &        $-0.04\pm0.16$                  &           $-0.06\pm0.08\pm0.01$       &          $-0.102\pm0.05\pm0.016$              &       $-0.09\pm0.04$
\end{tabular}
\end{ruledtabular}
\end{table}

\begin{table}[htb]
\caption{global fits of isospin amplitudes and strong phases with different $\gamma$s. 
The value of $\xi$ is fixed at 1.0}
\begin{center}
\begin{ruledtabular}
\begin{tabular}{ccccc}
& $\gamma=75^{\circ}$ & $\gamma=90^{\circ}$ & $\gamma=105^{\circ}$ & $\gamma=120^{\circ}$\\\hline
 $a^u_{1/2}$&$       467.8^{+        84.8}_{       -87.9}$&$          526.7^{+        93.3}_{      -100.1}$&$         589.6^{+       104.4}_{      -117.4}$&$       654.1^{+       116.6}_{      -141.2}$\\   
 $\d^u_{1/2}$&$         57.0^{+         0.3}_{        -0.2}$&$         63.3^{+         0.3}_{        -0.2}$&$            69.5^{+         0.3}_{        -0.2}$&$        69.5^{+         0.3}_{        -0.2}$\\  
 $a^c_{1/2}$&$       -112.7^{+        17.9}_{       -12.8}$&$       -107.9^{+        18.3}_{       -13.2}$&$         -104.3^{+        19.1}_{       -13.7}$&$      -103.4^{+        19.8}_{       -13.8}$\\   
 $a^u_{3/2}$&$        581.3^{+        48.2}_{       -52.2}$&$        631.0^{+        47.9}_{       -52.0}$&$          691.2^{+        50.9}_{       -55.3}$&$       745.0^{+        58.7}_{       -63.5}$\\   
 $\d^u_{3/2}$&$      -8066.2^{+         0.1}_{        -0.1}$&$       -180.8^{+         0.1}_{        -0.1}$&$          -218.5^{+         0.1}_{        -0.2}$&$      -331.6^{+         0.1}_{        -0.2}$\\ 
 $b^c_{1/2}$&$       -121.6^{+        16.8}_{       -12.7}$&$       -117.1^{+        17.3}_{       -13.4}$& $        -114.3^{+        18.3}_{       -14.1}$&$      -114.6^{+        19.6}_{       -14.6}$\\   
 $\d'^u_{1/2}$&$        213.4^{+         0.2}_{        -0.2}$&$        175.6^{+         0.2}_{        -0.3}$&$          225.9^{+         0.2}_{        -0.3}$&$      5183.4^{+         0.3}_{        -0.3}$\\
 $\k$&$            10.5^{+         2.9}_{        -2.8}$&$          9.8^{+         2.4}_{        -2.4}$&$              8.7^{+         2.0}_{        -2.1}$&$        7.4^{+         1.9}_{        -2.1}$\\       
$\chi^{2}_{min}$ & 0.80 & 0.77 & 0.79 & 0.83
\end{tabular}
\end{ruledtabular}
\end{center}
\label{gammas}
\end{table}
 
\begin{table}[htb]
\caption{Best fit values of isospin amlitudes with different value of
$\D^{u}_{1/2}$,  $\D^{c}_{1/2}$,$\D^{u}_{3/2}$,$\D^{c}_{3/2}$with gamma fixed at
$\fr{\pi}{3}(60^{\circ})$.
}\label{g60}
\begin{center}
\begin{ruledtabular}
\begin{tabular}{ccccc}
$\D^{u}_{1/2}$ & $-\pi/3$  & $-\pi/6$  & $+\pi/6$  & $+\pi/3$ \\\hline 
$ a^u_{1/2}$ &$       634.7^{+        97.0}_{      -114.5}$ & $       722.5^{+        89.4}_{      -104.3}$ & $       628.9^{+       102.4}_{      -105.2}$ & $       625.5^{+        96.0}_{       -98.7}$ \\
$\d^u_{1/2}$ &$         3.6^{+         0.2}_{        -0.3}$ & $         3.4^{+         0.2}_{        -0.2}$ & $         2.1^{+         0.3}_{        -0.2}$ & $         1.5^{+         0.3}_{        -0.3}$ \\
$ a^c_{1/2}$ &$      -112.7^{+        26.8}_{       -15.3}$ & $      -126.3^{+        12.2}_{        -6.0}$ & $      -135.3^{+         3.4}_{        -3.3}$ & $      -138.3^{+         3.6}_{        -3.6}$ \\
$ a^u_{3/2}$ &$       564.8^{+        69.8}_{       -74.6}$ & $       592.5^{+        63.1}_{       -69.2}$ & $       657.0^{+        48.2}_{       -52.3}$ & $       661.2^{+        46.8}_{       -50.3}$ \\
$\d^u_{3/2}$ &$         1.5^{+         0.2}_{        -0.2}$ & $         1.7^{+         0.3}_{        -0.2}$ & $         3.3^{+         0.3}_{        -0.3}$ & $         3.4^{+         0.3}_{        -0.3}$ \\
$ b^c_{1/2}$ &$      -129.5^{+        24.3}_{       -14.6}$ & $      -143.1^{+        11.8}_{        -6.2}$ & $      -141.1^{+         4.2}_{        -4.1}$ & $      -141.1^{+         4.2}_{        -4.1}$ \\
$\d'^u_{1/2}$ &$        -0.1^{+         0.4}_{        -0.4}$ & $         0.2^{+         0.5}_{        -0.3}$ & $         3.0^{+         0.4}_{        -0.4}$ & $         3.0^{+         0.4}_{        -0.4}$ \\
$        \k$ &$         9.8^{+         5.6}_{        -4.7}$ & $         5.3^{+         4.2}_{        -2.7}$ & $         1.6^{+         0.4}_{        -0.4}$ & $         1.5^{+         0.4}_{        -0.4}$ \\
$\chi^2_{min}$ &          2.3 &          3.3 &          4.1 &          5.2 \\\hline
$\D^{c}_{1/2}$ & $-\pi/3$  & $-\pi/6$  & $+\pi/6$  & $+\pi/3$ \\\hline 
$ a^u_{1/2}$ &$       686.1^{+       103.7}_{      -122.1}$ & $       668.0^{+       102.4}_{      -113.1}$ & $       581.8^{+        97.9}_{       -99.0}$ & $       503.2^{+        99.2}_{      -101.0}$ \\
$\d^u_{1/2}$ &$         2.1^{+         0.2}_{        -0.2}$ & $         2.3^{+         0.2}_{        -0.2}$ & $         2.5^{+         0.3}_{        -0.2}$ & $         2.5^{+         0.3}_{        -0.3}$ \\
$ a^c_{1/2}$ &$      -134.2^{+         3.5}_{        -3.5}$ & $      -134.0^{+         3.4}_{        -3.3}$ & $      -133.7^{+         3.3}_{        -3.3}$ & $      -134.4^{+         3.4}_{        -3.3}$ \\
$ a^u_{3/2}$ &$       651.1^{+        49.2}_{       -53.5}$ & $       651.1^{+        49.0}_{       -53.3}$ & $       651.6^{+        49.1}_{       -53.2}$ & $       650.6^{+        49.1}_{       -53.3}$ \\
$\d^u_{3/2}$ &$         3.1^{+         0.3}_{        -0.3}$ & $         3.2^{+         0.3}_{        -0.3}$ & $         3.2^{+         0.3}_{        -0.3}$ & $         3.2^{+         0.3}_{        -0.3}$ \\
$ b^c_{1/2}$ &$      -141.3^{+         4.2}_{        -4.1}$ & $      -141.3^{+         4.2}_{        -4.1}$ & $      -141.2^{+         4.2}_{        -4.1}$ & $      -141.2^{+         4.2}_{        -4.1}$ \\
$\d'^u_{1/2}$ &$         2.8^{+         0.4}_{        -0.5}$ & $         2.8^{+         0.4}_{        -0.4}$ & $         2.9^{+         0.4}_{        -0.4}$ & $         2.9^{+         0.4}_{        -0.4}$ \\
$        \k$ &$         1.5^{+         0.4}_{        -0.4}$ & $         1.5^{+         0.4}_{        -0.4}$ & $         1.6^{+         0.4}_{        -0.4}$ & $         1.6^{+         0.4}_{        -0.4}$ \\
$\chi^2_{min}$ &         16.5 &          7.7 &          2.4 &          3.2 \\\hline
$\D^{u}_{3/2}$ & $-\pi/3$  & $-\pi/6$  & $+\pi/6$  & $+\pi/3$ \\\hline 
$ a^u_{1/2}$ &$       547.8^{+       108.8}_{      -100.4}$ & $       594.8^{+       102.4}_{      -100.8}$ & $       670.1^{+        98.6}_{      -109.8}$ & $       701.1^{+        97.9}_{      -117.3}$ \\
$\d^u_{1/2}$ &$         2.0^{+         0.9}_{        -0.3}$ & $         2.2^{+         0.3}_{        -0.2}$ & $         2.6^{+         0.2}_{        -0.2}$ & $         2.8^{+         0.2}_{        -0.2}$ \\
$ a^c_{1/2}$ &$      -135.9^{+         3.5}_{        -3.5}$ & $      -134.7^{+         3.4}_{        -3.3}$ & $      -132.7^{+         3.3}_{        -3.2}$ & $      -131.7^{+         3.3}_{        -3.2}$ \\
$ a^u_{3/2}$ &$       661.9^{+        49.4}_{       -53.3}$ & $       656.0^{+        48.9}_{       -53.1}$ & $       648.9^{+        49.4}_{       -53.5}$ & $       648.2^{+        49.9}_{       -54.2}$ \\
$\d^u_{3/2}$ &$         3.6^{+         0.4}_{        -0.7}$ & $         3.4^{+         0.3}_{        -0.3}$ & $         3.0^{+         0.2}_{        -0.3}$ & $         2.7^{+         0.2}_{        -0.3}$ \\
$ b^c_{1/2}$ &$      -141.0^{+         4.2}_{        -4.1}$ & $      -141.1^{+         4.2}_{        -4.1}$ & $      -141.5^{+         4.3}_{        -4.2}$ & $      -142.1^{+         4.3}_{        -4.3}$ \\
$\d'^u_{1/2}$ &$         3.2^{+         0.4}_{        -0.4}$ & $         3.1^{+         0.4}_{        -0.4}$ & $         2.7^{+         0.4}_{        -0.5}$ & $         2.4^{+         0.4}_{        -0.6}$ \\
$        \k$ &$         1.6^{+         0.4}_{        -0.4}$ & $         1.6^{+         0.4}_{        -0.4}$ & $         1.6^{+         0.4}_{        -0.4}$ & $         1.6^{+         0.4}_{        -0.4}$ \\
$\chi^2_{min}$ &          2.7 &          2.3 &          4.9 &          7.9 \\\hline
$\D^{c}_{3/2}$ & $-\pi/3$  & $-\pi/6$  & $+\pi/6$  & $+\pi/3$ \\\hline 
$ a^u_{1/2}$ &$       632.0^{+       100.6}_{      -106.1}$ & $       632.9^{+       100.4}_{      -105.3}$ & $       634.3^{+        99.9}_{      -104.5}$ & $       634.8^{+        99.5}_{      -104.6}$ \\
$\d^u_{1/2}$ &$         2.4^{+         0.2}_{        -0.2}$ & $         2.4^{+         0.2}_{        -0.2}$ & $         2.4^{+         0.2}_{        -0.2}$ & $         2.4^{+         0.2}_{        -0.2}$ \\
$ a^c_{1/2}$ &$      -133.7^{+         3.3}_{        -3.3}$ & $      -133.7^{+         3.3}_{        -3.3}$ & $      -133.6^{+         3.3}_{        -3.3}$ & $      -133.5^{+         3.3}_{        -3.3}$ \\
$ a^u_{3/2}$ &$       659.9^{+        49.0}_{       -53.2}$ & $       653.9^{+        49.0}_{       -53.2}$ & $       653.9^{+        48.9}_{       -53.1}$ & $       659.6^{+        48.9}_{       -53.0}$ \\
$\d^u_{3/2}$ &$         3.2^{+         0.2}_{        -0.3}$ & $         3.2^{+         0.2}_{        -0.3}$ & $         3.2^{+         0.3}_{        -0.3}$ & $         3.2^{+         0.2}_{        -0.3}$ \\
$ b^c_{1/2}$ &$      -141.2^{+         4.2}_{        -4.1}$ & $      -141.2^{+         4.2}_{        -4.1}$ & $      -141.2^{+         4.2}_{        -4.1}$ & $      -141.2^{+         4.2}_{        -4.1}$ \\
$\d'^u_{1/2}$ &$         2.9^{+         0.4}_{        -0.4}$ & $         2.9^{+         0.4}_{        -0.4}$ & $         2.9^{+         0.4}_{        -0.4}$ & $         2.9^{+         0.4}_{        -0.4}$ \\
$        \k$ &$         1.6^{+         0.4}_{        -0.3}$ & $         1.6^{+         0.4}_{        -0.4}$ & $         1.5^{+         0.4}_{        -0.4}$ & $         1.5^{+         0.4}_{        -0.3}$ \\
$\chi^2_{min}$ &          2.5 &          2.6 &          3.4 &          3.8 \\
\end{tabular}
\end{ruledtabular}
\end{center}
\end{table}

\begin{table}[htb]
\caption{
Best fit values of isospin amlitudes with different value of
$\D^{u}_{1/2}$,  $\D^{c}_{1/2}$,$\D^{u}_{3/2}$,$\D^{c}_{3/2}$with gamma fixed at
$\fr{\pi}{2}(90^{\circ})$.
}\label{g90}
\begin{center}
\begin{ruledtabular}
\begin{tabular}{ccccc}
$\D^{u}_{1/2}$ & $-\pi/3$  & $-\pi/6$  & $+\pi/6$  & $+\pi/3$ \\\hline 
$ a^u_{1/2}$ &$       567.4^{+       103.0}_{      -108.2}$ & $       515.1^{+        89.7}_{       -88.2}$ & $       604.4^{+        80.0}_{       -87.0}$ & $       638.9^{+        87.2}_{      -312.4}$ \\
$\d^u_{1/2}$ &$         1.8^{+         0.6}_{        -0.5}$ & $         1.0^{+         0.5}_{        -0.4}$ & $         0.4^{+         0.4}_{        -0.3}$ & $         0.4^{+         0.6}_{        -0.4}$ \\
$ a^c_{1/2}$ &$       -42.0^{+         4.7}_{       -11.9}$ & $       -42.6^{+         5.7}_{       -15.0}$ & $       -39.2^{+         3.8}_{        -6.8}$ & $       -40.2^{+         4.7}_{        -0.0}$ \\
$ a^u_{3/2}$ &$       590.2^{+        53.2}_{       -58.9}$ & $       595.4^{+        52.2}_{       -57.6}$ & $       573.2^{+        55.7}_{       -61.9}$ & $       574.3^{+       142.9}_{       -64.5}$ \\
$\d^u_{3/2}$ &$        -0.3^{+         0.5}_{        -0.5}$ & $        -0.4^{+         0.5}_{        -0.5}$ & $        -0.2^{+         0.5}_{        -0.4}$ & $         0.3^{+         1.4}_{        -0.5}$ \\
$ b^c_{1/2}$ &$       -60.8^{+        14.3}_{       -12.9}$ & $       -60.5^{+        14.7}_{       -14.8}$ & $       -60.2^{+        14.0}_{       -12.1}$ & $       -64.0^{+        14.3}_{       -78.7}$ \\
$\d'^u_{1/2}$ &$        -2.5^{+         0.6}_{        -0.5}$ & $        -2.6^{+         0.6}_{        -0.4}$ & $        -2.4^{+         0.6}_{        -0.5}$ & $        -1.9^{+         2.2}_{        -0.6}$ \\
$        \k$ &$        16.1^{+         1.9}_{        -1.5}$ & $        16.0^{+         1.8}_{        -1.4}$ & $        16.7^{+         2.1}_{        -1.6}$ & $        16.5^{+         2.3}_{       -12.9}$ \\
$\chi^2_{min}$ &          3.1 &          3.4 &          1.3 &          3.1 \\\hline
$\D^{c}_{1/2}$ & $-\pi/3$  & $-\pi/6$  & $+\pi/6$  & $+\pi/3$ \\\hline 
$ a^u_{1/2}$ &$       541.5^{+        79.6}_{       -79.3}$ & $       544.6^{+        83.5}_{       -83.3}$ & $       559.8^{+        82.0}_{       -85.3}$ & $       571.5^{+        79.0}_{       -82.8}$ \\
$\d^u_{1/2}$ &$         0.5^{+         0.4}_{        -0.3}$ & $         0.5^{+         0.4}_{        -0.3}$ & $         0.6^{+         0.4}_{        -0.4}$ & $         0.7^{+         0.4}_{        -0.4}$ \\
$ a^c_{1/2}$ &$       -41.2^{+         5.0}_{       -13.2}$ & $       -41.6^{+         5.3}_{       -14.2}$ & $       -40.4^{+         4.4}_{        -9.2}$ & $       -39.8^{+         4.2}_{        -8.0}$ \\
$ a^u_{3/2}$ &$       586.2^{+        53.8}_{       -59.8}$ & $       585.4^{+        53.9}_{       -59.9}$ & $       581.3^{+        54.6}_{       -60.7}$ & $       578.4^{+        54.8}_{       -61.0}$ \\
$\d^u_{3/2}$ &$        -0.4^{+         0.5}_{        -0.5}$ & $        -0.4^{+         0.5}_{        -0.5}$ & $        -0.3^{+         0.5}_{        -0.4}$ & $        -0.3^{+         0.5}_{        -0.4}$ \\
$ b^c_{1/2}$ &$       -60.2^{+        14.4}_{       -13.6}$ & $       -60.3^{+        14.5}_{       -14.1}$ & $       -59.7^{+        14.2}_{       -12.5}$ & $       -59.2^{+        14.1}_{       -12.4}$ \\
$\d'^u_{1/2}$ &$        -2.6^{+         0.6}_{        -0.5}$ & $        -2.6^{+         0.6}_{        -0.5}$ & $        -2.5^{+         0.6}_{        -0.5}$ & $        -2.5^{+         0.5}_{        -0.5}$ \\
$        \k$ &$        16.3^{+         2.0}_{        -1.5}$ & $        16.3^{+         2.0}_{        -1.5}$ & $        16.4^{+         2.0}_{        -1.5}$ & $        16.6^{+         2.1}_{        -1.6}$ \\
$\chi^2_{min}$ &          1.5 &          1.6 &          2.8 &          3.8 \\\hline
$\D^{u}_{3/2}$ & $-\pi/3$  & $-\pi/6$  & $+\pi/6$  & $+\pi/3$ \\\hline 
$ a^u_{1/2}$ &$       478.7^{+       112.6}_{      -127.4}$ & $       486.9^{+       108.3}_{      -119.6}$ & $       528.2^{+       102.3}_{      -117.1}$ & $       555.2^{+        99.5}_{      -118.9}$ \\
$\d^u_{1/2}$ &$         1.8^{+         1.2}_{        -0.3}$ & $         2.1^{+         0.4}_{        -0.2}$ & $         2.5^{+         0.2}_{        -0.2}$ & $         2.8^{+         0.2}_{        -0.2}$ \\
$ a^c_{1/2}$ &$      -137.7^{+         3.2}_{        -3.1}$ & $      -137.8^{+         3.2}_{        -3.1}$ & $      -137.7^{+         3.2}_{        -3.1}$ & $      -137.4^{+         3.2}_{        -3.1}$ \\
$ a^u_{3/2}$ &$       667.1^{+        49.2}_{       -52.9}$ & $       667.0^{+        48.5}_{       -52.7}$ & $       656.9^{+        49.4}_{       -53.7}$ & $       647.8^{+        50.2}_{       -54.6}$ \\
$\d^u_{3/2}$ &$         3.6^{+         0.4}_{        -0.7}$ & $         3.5^{+         0.3}_{        -0.4}$ & $         3.0^{+         0.2}_{        -0.3}$ & $         2.8^{+         0.2}_{        -0.3}$ \\
$ b^c_{1/2}$ &$      -144.3^{+         4.2}_{        -4.1}$ & $      -144.3^{+         4.2}_{        -4.1}$ & $      -144.3^{+         4.2}_{        -4.1}$ & $      -144.4^{+         4.2}_{        -4.1}$ \\
$\d'^u_{1/2}$ &$         3.2^{+         0.4}_{        -0.7}$ & $         3.1^{+         0.4}_{        -0.4}$ & $         2.7^{+         0.3}_{        -0.4}$ & $         2.5^{+         0.4}_{        -0.5}$ \\
$        \k$ &$         0.9^{+         0.4}_{        -0.4}$ & $         0.9^{+         0.4}_{        -0.4}$ & $         1.0^{+         0.4}_{        -0.3}$ & $         1.0^{+         0.4}_{        -0.4}$ \\
$\chi^2_{min}$ &          2.7 &          2.1 &          5.3 &          9.1 \\\hline
$\D^{c}_{3/2}$ & $-\pi/3$  & $-\pi/6$  & $+\pi/6$  & $+\pi/3$ \\\hline 
$ a^u_{1/2}$ &$       632.0^{+       100.6}_{      -106.1}$ & $       632.9^{+       100.4}_{      -105.3}$ & $       634.3^{+        99.9}_{      -104.5}$ & $       634.8^{+        99.5}_{      -104.6}$ \\
$\d^u_{1/2}$ &$         2.4^{+         0.2}_{        -0.2}$ & $         2.4^{+         0.2}_{        -0.2}$ & $         2.4^{+         0.2}_{        -0.2}$ & $         2.4^{+         0.2}_{        -0.2}$ \\
$ a^c_{1/2}$ &$      -133.7^{+         3.3}_{        -3.3}$ & $      -133.7^{+         3.3}_{        -3.3}$ & $      -133.6^{+         3.3}_{        -3.3}$ & $      -133.5^{+         3.3}_{        -3.3}$ \\
$ a^u_{3/2}$ &$       659.9^{+        49.0}_{       -53.2}$ & $       653.9^{+        49.0}_{       -53.2}$ & $       653.9^{+        48.9}_{       -53.1}$ & $       659.6^{+        48.9}_{       -53.0}$ \\
$\d^u_{3/2}$ &$         3.2^{+         0.2}_{        -0.3}$ & $         3.2^{+         0.2}_{        -0.3}$ & $         3.2^{+         0.3}_{        -0.3}$ & $         3.2^{+         0.2}_{        -0.3}$ \\
$ b^c_{1/2}$ &$      -141.2^{+         4.2}_{        -4.1}$ & $      -141.2^{+         4.2}_{        -4.1}$ & $      -141.2^{+         4.2}_{        -4.1}$ & $      -141.2^{+         4.2}_{        -4.1}$ \\
$\d'^u_{1/2}$ &$         2.9^{+         0.4}_{        -0.4}$ & $         2.9^{+         0.4}_{        -0.4}$ & $         2.9^{+         0.4}_{        -0.4}$ & $         2.9^{+         0.4}_{        -0.4}$ \\
$        \k$ &$         1.6^{+         0.4}_{        -0.3}$ & $         1.6^{+         0.4}_{        -0.4}$ & $         1.5^{+         0.4}_{        -0.4}$ & $         1.5^{+         0.4}_{        -0.3}$ \\
$\chi^2_{min}$ &          2.5 &          2.6 &          3.4 &          3.8 \\
\end{tabular}
\end{ruledtabular}
\end{center}
\end{table}

\begin{table}[htb]
\caption{Best fit values of isospin amlitudes with different value of
$\D^{u}_{1/2}$,  $\D^{c}_{1/2}$,$\D^{u}_{3/2}$,$\D^{c}_{3/2}$with gamma fixed at
$\fr{2\pi}{3}(120^{\circ})$.}\label{g120}
\begin{center}
\begin{ruledtabular}
\begin{tabular}{ccccc}
$\D^{u}_{1/2}$ & $-\pi/3$  & $-\pi/6$  & $+\pi/6$  & $+\pi/3$ \\\hline 
$ a^u_{1/2}$ &$       440.7^{+       127.5}_{      -118.0}$ & $       459.6^{+       110.4}_{      -140.6}$ & $       434.6^{+       113.6}_{      -144.0}$ & $       471.2^{+       132.6}_{      -156.4}$ \\
$\d^u_{1/2}$ &$         3.6^{+         0.3}_{        -0.3}$ & $         2.3^{+         0.2}_{        -0.2}$ & $         1.4^{+         0.3}_{        -0.3}$ & $         0.8^{+         0.4}_{        -0.4}$ \\
$ a^c_{1/2}$ &$      -133.1^{+        41.6}_{       -13.1}$ & $      -141.2^{+         3.3}_{        -3.2}$ & $      -137.7^{+         3.4}_{        -3.3}$ & $      -134.9^{+         3.8}_{        -3.7}$ \\
$ a^u_{3/2}$ &$       693.2^{+        74.6}_{       -58.1}$ & $       668.2^{+        48.7}_{       -52.8}$ & $       670.7^{+        48.2}_{       -52.1}$ & $       668.0^{+        48.8}_{       -52.7}$ \\
$\d^u_{3/2}$ &$         1.6^{+         0.3}_{        -0.4}$ & $         3.2^{+         0.2}_{        -0.3}$ & $         3.3^{+         0.3}_{        -0.3}$ & $         3.1^{+         0.3}_{        -0.4}$ \\
$ b^c_{1/2}$ &$      -132.6^{+        38.5}_{       -20.5}$ & $      -148.3^{+         4.2}_{        -4.1}$ & $      -148.4^{+         4.2}_{        -4.1}$ & $      -148.3^{+         4.2}_{        -4.1}$ \\
$\d'^u_{1/2}$ &$         0.1^{+         0.8}_{        -0.7}$ & $         2.8^{+         0.4}_{        -0.4}$ & $         2.9^{+         0.4}_{        -0.4}$ & $         2.8^{+         0.4}_{        -0.5}$ \\
$        \k$ &$         4.3^{+         5.4}_{        -3.4}$ & $         0.2^{+         0.3}_{        -0.3}$ & $         0.2^{+         0.3}_{        -0.3}$ & $         0.2^{+         0.3}_{        -0.3}$ \\
$\chi^2_{min}$ &          3.3 &          2.0 &          2.3 &          2.0 \\\hline
$\D^{c}_{1/2}$ & $-\pi/3$  & $-\pi/6$  & $+\pi/6$  & $+\pi/3$ \\\hline 
$ a^u_{1/2}$ &$       564.5^{+       116.6}_{      -135.2}$ & $       622.6^{+       119.0}_{      -142.3}$ & $       623.3^{+       122.0}_{      -150.3}$ & $       629.2^{+        84.5}_{       -87.9}$ \\
$\d^u_{1/2}$ &$         0.3^{+         0.4}_{        -0.2}$ & $         0.3^{+         0.3}_{        -0.2}$ & $         0.6^{+         0.4}_{        -0.2}$ & $         0.8^{+         0.5}_{        -0.4}$ \\
$ a^c_{1/2}$ &$      -103.6^{+        21.3}_{       -15.6}$ & $      -102.8^{+        20.1}_{       -14.5}$ & $      -100.4^{+        26.2}_{       -15.6}$ & $       -35.7^{+         3.9}_{        -7.2}$ \\
$ a^u_{3/2}$ &$       744.1^{+        58.6}_{       -63.7}$ & $       746.1^{+        58.4}_{       -63.3}$ & $       759.2^{+        59.2}_{       -63.7}$ & $       770.7^{+        52.2}_{       -57.6}$ \\
$\d^u_{3/2}$ &$         1.4^{+         0.1}_{        -0.2}$ & $         1.4^{+         0.1}_{        -0.2}$ & $         1.4^{+         0.1}_{        -0.2}$ & $        -0.2^{+         0.5}_{        -0.5}$ \\
$ b^c_{1/2}$ &$      -114.2^{+        20.5}_{       -15.8}$ & $      -113.9^{+        19.7}_{       -15.1}$ & $      -111.1^{+        24.4}_{       -16.2}$ & $       -62.9^{+        13.1}_{       -12.1}$ \\
$\d'^u_{1/2}$ &$        -0.3^{+         0.3}_{        -0.3}$ & $        -0.3^{+         0.3}_{        -0.3}$ & $        -0.3^{+         0.3}_{        -0.4}$ & $        -2.5^{+         0.6}_{        -0.5}$ \\
$        \k$ &$         7.4^{+         2.0}_{        -2.4}$ & $         7.5^{+         1.9}_{        -2.2}$ & $         7.7^{+         2.0}_{        -2.1}$ & $        11.7^{+         1.1}_{        -0.9}$ \\
$\chi^2_{min}$ &          1.4 &          1.3 &          4.2 &          3.6 \\\hline
$\D^{u}_{3/2}$ & $-\pi/3$  & $-\pi/6$  & $+\pi/6$  & $+\pi/3$ \\\hline 
$ a^u_{1/2}$ &$       530.5^{+       124.3}_{      -186.7}$ & $       473.1^{+       113.1}_{      -154.9}$ & $       414.5^{+       105.7}_{      -133.1}$ & $       403.5^{+       104.3}_{      -139.0}$ \\
$\d^u_{1/2}$ &$         1.4^{+         0.3}_{        -0.3}$ & $         1.7^{+         0.2}_{        -0.2}$ & $         2.2^{+         0.2}_{        -0.2}$ & $         2.5^{+         0.3}_{        -0.2}$ \\
$ a^c_{1/2}$ &$      -136.7^{+         3.6}_{        -3.4}$ & $      -138.5^{+         3.3}_{        -3.2}$ & $      -140.6^{+         3.3}_{        -3.2}$ & $      -141.1^{+         3.3}_{        -3.2}$ \\
$ a^u_{3/2}$ &$       669.3^{+        49.0}_{       -52.7}$ & $       671.1^{+        48.4}_{       -52.4}$ & $       663.6^{+        49.3}_{       -53.4}$ & $       655.8^{+        50.3}_{       -54.5}$ \\
$\d^u_{3/2}$ &$         3.7^{+         0.4}_{        -0.4}$ & $         3.5^{+         0.3}_{        -0.3}$ & $         3.0^{+         0.2}_{        -0.3}$ & $         2.7^{+         0.2}_{        -0.4}$ \\
$ b^c_{1/2}$ &$      -148.4^{+         4.2}_{        -4.1}$ & $      -148.4^{+         4.2}_{        -4.1}$ & $      -148.1^{+         4.3}_{        -4.1}$ & $      -147.5^{+         4.4}_{        -4.2}$ \\
$\d'^u_{1/2}$ &$         3.2^{+         0.4}_{        -0.4}$ & $         3.1^{+         0.4}_{        -0.4}$ & $         2.7^{+         0.4}_{        -0.5}$ & $         2.4^{+         0.4}_{        -0.7}$ \\
$        \k$ &$         0.2^{+         0.4}_{        -0.4}$ & $         0.2^{+         0.4}_{        -0.4}$ & $         0.2^{+         0.3}_{        -0.3}$ & $         0.3^{+         0.4}_{        -0.4}$ \\
$\chi^2_{min}$ &          1.5 &          1.3 &          4.5 &          8.0 \\\hline
$\D^{c}_{3/2}$ & $-\pi/3$  & $-\pi/6$  & $+\pi/6$  & $+\pi/3$ \\\hline 
$ a^u_{1/2}$ &$       435.6^{+       108.7}_{      -140.3}$ & $       435.5^{+       108.9}_{      -140.5}$ & $       435.6^{+       109.0}_{      -140.5}$ & $       435.7^{+       109.0}_{      -140.3}$ \\
$\d^u_{1/2}$ &$         1.9^{+         0.2}_{        -0.2}$ & $         1.9^{+         0.2}_{        -0.2}$ & $         1.9^{+         0.2}_{        -0.2}$ & $         1.9^{+         0.2}_{        -0.2}$ \\
$ a^c_{1/2}$ &$      -139.8^{+         3.3}_{        -3.2}$ & $      -139.8^{+         3.3}_{        -3.2}$ & $      -139.7^{+         3.3}_{        -3.2}$ & $      -139.6^{+         3.3}_{        -3.2}$ \\
$ a^u_{3/2}$ &$       667.6^{+        48.9}_{       -53.1}$ & $       668.7^{+        48.7}_{       -52.8}$ & $       669.1^{+        48.6}_{       -52.7}$ & $       668.6^{+        48.8}_{       -52.8}$ \\
$\d^u_{3/2}$ &$         3.3^{+         0.3}_{        -0.3}$ & $         3.3^{+         0.3}_{        -0.3}$ & $         3.3^{+         0.3}_{        -0.3}$ & $         3.3^{+         0.3}_{        -0.3}$ \\
$ b^c_{1/2}$ &$      -148.3^{+         4.2}_{        -4.1}$ & $      -148.3^{+         4.2}_{        -4.1}$ & $      -148.4^{+         4.2}_{        -4.1}$ & $      -148.4^{+         4.2}_{        -4.1}$ \\
$\d'^u_{1/2}$ &$         2.9^{+         0.4}_{        -0.4}$ & $         2.9^{+         0.4}_{        -0.4}$ & $         2.9^{+         0.4}_{        -0.4}$ & $         2.9^{+         0.4}_{        -0.4}$ \\
$        \k$ &$         0.2^{+         0.4}_{        -0.4}$ & $         0.2^{+         0.3}_{        -0.3}$ & $         0.2^{+         0.3}_{        -0.3}$ & $         0.2^{+         0.3}_{        -0.3}$ \\
$\chi^2_{min}$ &          2.2 &          2.2 &          2.2 &          2.2 \\
\end{tabular}
\end{ruledtabular}
\end{center}
\end{table}
\end{document}